\documentclass[a4paper,fleqn,usenatbib]{mnras}

\usepackage{mathptmx}
\usepackage{txfonts}
\usepackage[T1]{fontenc}
\usepackage{ae,aecompl}
\usepackage{multicol}
\usepackage{graphicx}
\usepackage{color}
\usepackage{graphicx}
\usepackage{amssymb}
\usepackage{rotating}
\usepackage{longtable}
\usepackage{lscape}
\usepackage{multirow}
\usepackage{multicol}

\newcommand{\figref}{Figure~\ref}

\title[Extratidal features in the globular cluster NGC\,362]{Using Gaia DR2 to detect extratidal structures around the Galactic globular cluster NGC\,362}

\author[J. A. Carballo-Bello et al.]{Julio A. Carballo-Bello$^{1,2}$\thanks{E-mail: jcarballo@astro.puc.cl}\\
$^{1}$Instituto de Astrof\'isica, Facultad de F\'isica, Pontificia Universidad Cat\'olica de Chile, Av. Vicu\~na Mackenna 4860, 782-0436 Macul,\\ 
Santiago, Chile\\
$^{2}$Chinese Academy of Sciences South America  Center for Astronomy, National Astronomical Observatories, CAS, Beijing 100101, China
}

\date{Accepted XXX. Received YYY; in original form ZZZ}

\pubyear{2018}

\begin{document}
\label{firstpage}
\pagerange{\pageref{firstpage}--\pageref{lastpage}}
\maketitle

\begin{abstract}
We explore the possibility of searching for extratidal features around the Galactic globular cluster NGC\,362 using the {\it Gaia} DR2 together with a modified version of a classical statistical decontamination algorithm. Our results suggest that an important stellar component is associated with this globular cluster, which is perfectly distinguishable from the populations in the Small Magellanic Cloud and that of the nearby NGC\,104 (47\,Tucanae). We thus confirm once again the power of {\it Gaia} to disentangle different stellar components along the same line-of-sight.  
\end{abstract}

\begin{keywords}
(Galaxy): globular clusters: individual: NGC\,362 (Galaxy)- halo
\end{keywords}

\section{Introduction}

The tidal stress exerted by the Milky Way on Galactic globular clusters (GCs) has a direct impact on their structure and evolution and varies in time as these systems move along their orbits. Indeed, the successive passages through the disk and at pericenters contribute to the loss of high energy stars via tidal stripping \citep[e.g.][]{Combes1999,Johnston1999}. The result of that mechanism is the formation of more or less extended tidal tails, which have been succesfully traced across the sky for a handful of globulars in the Galaxy. The tidal tails generated by the disruption of Pal\,5 \citep{Odenkirchen2003,Grillmair2006a,Kuzma2015} and NGC\,5466 \citep{Belokurov2006,Grillmair2006} are well-studied examples of that phenomenon. On the other hand, for a more numerous group of GCs, those stars energetically unbound beyond the \cite{King1962} tidal radius but still found within the Jacobi surface of the cluster generate an `extended halo', which seems to be a common feature for many outer globulars \citep[e.g.][]{Olszewski2009,Correnti2011,Kuzma2016,Carballo-Bello2018}.

These extratidal structures have been typically unveiled thanks to different procedures that consider not only the color-magnitude distribution of the cluster stars but also that of the populations in the surrounding area. For example, the matched-filter technique has been profusely used to reveal the structures generated by the tidal disruption and evolution of GCs (and satellite galaxies) in the Milky Way \citep[e.g.][]{Odenkirchen2001,Odenkirchen2003,Balbinot2011,Sollima2011,Carballo-Bello2017,Navarrete2017}. As for clusters in the densest areas of the Galaxy, i.e. disk and bulge, the identification of the intrinsic color-magnitude diagram (CMD) features associated with the GC, specially in their low-density outer layers, requires an alternative approach. Statistical decontamination algorithms \citep[e.g. see][and references therein]{Bonatto2007}, designed to identify probable cluster members, have proven to be effective in the removal of polluting fore/background stars during the study of Galactic globular and open clusters \citep[see][]{Bonatto2008,Kurtev2008,Carballo-Bello2016}. Spectroscopic tagging of stars far from the GC centers using the information derived by wide-sky surveys is also a promising technique \citep[e.g.][]{Anguiano2015}.

Among the available datasets, the second data release (DR2) generated by the European Space Agency Mission {\it Gaia} \citep{Gaia2016,Gaia2018a} is providing insights on the origin, dynamical history and current structure of the Milky Way with unprecedented quality. As for the Galactic GCs, the proper motions derived by {\it Gaia} for relatively bright cluster members together with their mean line-of-sight velocities have allowed us to compute their orbits \citep{Gaia2018b,Baumgardt2019,Vasiliev2019}. Additional examples of recent results on GCs based on {\it Gaia} DR2 are the detection of tidal tails around \emph{w}\,Cen \citep{Ibata2019}, the obtaining of radial number density profiles for around half of the globulars in the Milky Way within their Jacobi radii  \citep{deBoer2019} and the detection of internal rotation in nearby clusters \citep{Bianchini2018,Sollima2019}. 

In this Paper, we propose to search for extratidal overdensities using a modified version of a statistical decontamination procedure in combination with the stellar parameters provided by {\it Gaia} DR2. One of the objectives of this work is to evaluate whether this technique is adequate for data with the current quality of the information contained in the {\it Gaia} DR2 or it will be (only) possible to identify extratidal features in nearby GCs using the end-of-mission data release.  
Among the nearby Galactic GCs, we focus on NGC\,362, a system at $R_{\odot} = 8.5$\,kpc \citep{Chen2018} with coordinates ($\alpha,\delta$)=(15.8099$^{\circ}$,-70.8489$^{\circ}$) \citep{Gaia2018b}. This cluster is an ideal target to test the method proposed in this Paper because $i)$ it is close enough to have a decent number of main-sequence (MS) stars observed below its Turn-Off (TO), $ii)$ its population is likely contaminated by Small Magellanic Cloud (SMC) stars, and $iii)$ a remarkable GC, NGC\,104 (47\,Tucanae), is located at only a few degrees away and it may also contribute stars in the field around NGC\,362. Interestingly, the proper motions derived for NGC\,104 and NGC\,362 are almost identical \citep{Gaia2018b}.

\section{Data}
\label{data}

  \begin{figure*}
     \begin{center}
      \includegraphics[scale=0.5]{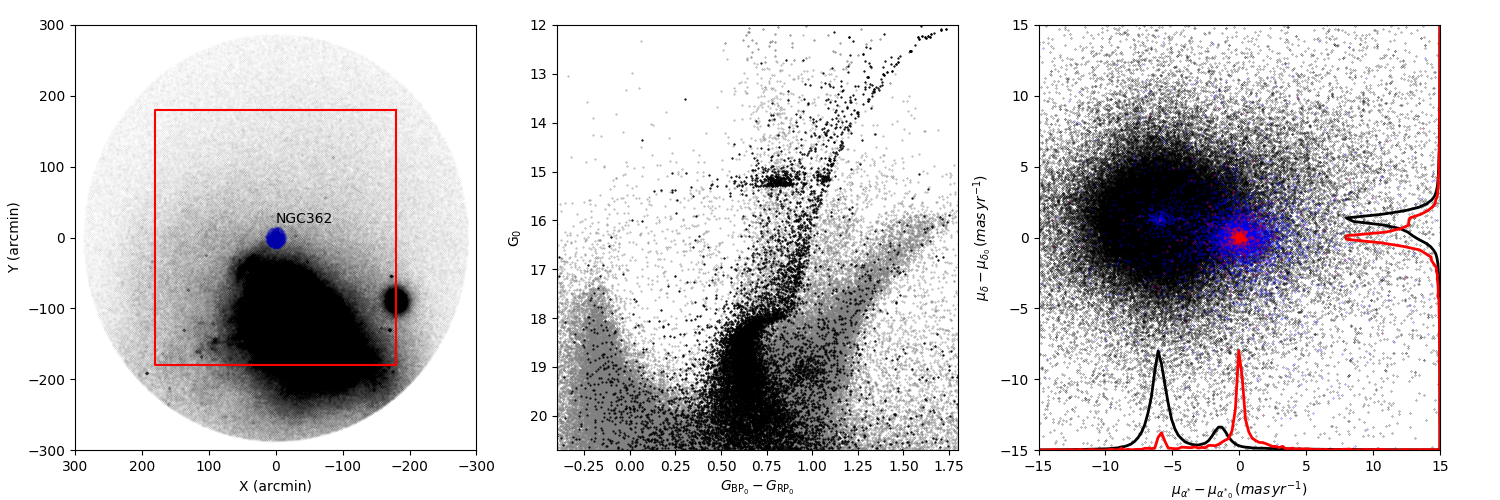}
      \caption[]{\emph{Left:} {\it Gaia} DR2 sources considered in this work around NGC\,362 (in blue). The SMC is located to the south of the GC while NGC\,104 is clearly visible at $\sim$\,3\,deg from NGC\,362. The separation between the areas used for analysis and as control field is indicated by the solid red line. \emph{Center:} CMD for stars within 15\,arcmin from the cluster center (black) and those stars in the range $60 < r < 90$\,arcmin (grey). The foreground population is almost entirely associated with the SMC. \emph{Right:} Gaia DR2 proper motion distribution for the same populations as in the middle panel, where black, blue and red points correspond to background, faint ($G_{0} > 18$) and bright ($G_{0} \leq 18$) MS NGC\,362 stars, respectively. The black and red solid lines indicate the histograms corresponding to the background (black) and bright stars in the cluster (red) populations.} 
\label{figura1}
     \end{center}
   \end{figure*}

The {\it Gaia} second data release (DR2) contains, among other parameters, trigonometric parallaxes, proper motions and three broad-band magnitudes ($G, G_{\rm BP}, G_{\rm RP}$) for more than a billion  objects \citep{Gaia2018a}. However, despite the remarkable quality of the data, the {\it Gaia} DR2 data present technical limitations and systematic uncertainties that make the parametrization of individual stars in crowded fields harder \citep[see e.g.][]{Pancino2017,Arenou2018,Evans2018}. While the $G$ magnitudes are less affected by crowding, the $G_{\rm BP}/G_{\rm RP}$ photometry for faint stars is severely altered due to the presence of unprocessed blends in the windows of the $BP$ and $RP$ photometers. Thus, the flux of a given target in those bands may be polluted by that of a nearby star and errors dramatically increase beyond $G \sim 18$.

These characteristics of the {\it Gaia} DR2 impose constraints on the study of extratidal populations around Galactic GCs. Given that those structures are mainly composed of faint (low-mass) stars, the detection of low surface-brightness features such as tidal tails and extended haloes is not possible for GCs at large heliocentric distances. Indeed, {\it Gaia} DR2 data contain a marginal (or zero)  number of stars fainter that $G \sim 21$, sampling only evolved stars in clusters beyond $R_{\odot} \sim$ 12--15\,kpc. Therefore, we should focus on closer clusters when searching for any stellar feature in their surroundings with {\it Gaia} DR2. 

We have retrieved all the information available in {\it Gaia} DR2 for objects within a box of 10\,deg centered on NGC\,362. The position of the globular with respect to NGC\,104 and the SMC system is shown in \figref{figura1}, where celestial positions are converted to cartesian coordinates with respect to the cluster center (X,Y) using the set of expressions proposed by \cite{vandeven2006}. The likely presence of SMC stars in the surroundings of NGC\,362 is evident in the CMD contained in the middle panel of the same figure. For a sample of stars with $r < 15$\,arcmin from the cluster center, the pollution in the diagram by members of the background galaxy is observed in the faintest section of the CMD. Fortunately, {\it Gaia} DR2 provides additional information that might allow us to distinguish between the SMC and  Milky Way fore/background stars and those stars belonging to the target GC. To illustrate this, we have obtained the distribution of $\mu_{\alpha^{*}}$ and $\mu_{\delta}$ both for stars likely belonging to the cluster ($r < 15$\,arcmin from its center) and for a sample of fore/background objects ($60 < r < 90$\,arcmin) using a bin size of 0.1\,mas\,yr$^{-1}$. As shown in the right panel in \figref{figura1}, the distribution of $\mu_{\delta}$ values is similar for both populations. On the other hand, the $\mu_{\alpha^{*}}$ distribution shows that the motion of NGC\,362 is significantly different from that of the bulk of fore/background stars. We then conclude that this GC is the perfect target to test our method for the detection of stellar overdensities based on its CMD morphology and proper motions. As expected, if we split the cluster population into two groups with $G_{0} \leq 18$ and $G_{0} > 18$, the latter presents a more dispersed distribution as shown in \figref{figura1}. The adopted values for $\mu_{\alpha^{*}}$ and $\mu_{\delta}$ for this work thus correspond to the peak value in the distribution for stars with $r < 10$\,arcmin and $G_{0} < 18$, resulting in ($\mu_{\alpha^{*}},\mu_{\delta}$) = (-6.7$\pm$0.3, -2.5$\pm$ 0.8) in mas\,yr$^{-1}$. These values are in good agreement with those derived by \cite{Gaia2018b} and \cite{Libralato2018}.  

$E(B-V)$ values have been derived for each object from \cite{Schlegel1998} dust maps and we have adopted the {\it Gaia} extinction coefficients provided by \cite{Gaia2018a}. The photometric errors were computed using $\sigma_{\rm mag}^{2} = (1.086\sigma_{\rm flux}/{\rm flux})^{2} + (\sigma_{\rm zp})^{2}$, where $\sigma_{\rm flux}$ and $\sigma_{\rm zp}$ are the error in the flux and photometric zeropoint, respectively \citep{Evans2018}. We make use of the parameter \textsc{phot\_bp\_rp\_excess\_factor}, which indicates the inconsistency between the sum of the fluxes in $G_{\rm BP}$ and $G_{\rm RP}$ with respect to the white light $G$ broad band. We limit our final catalogs to stars with \textsc{phot\_bp\_rp\_excess\_factor} < 1.5 and only consider those objects with at least 5 independent scans by {\it Gaia} (\textsc{visibility\_periods\_used} $\geq$ 5).

\section{Methodology}

The algorithm designed for this work is based on the statistical decontamination procedure described in \citealt[][and references therein]{Bonatto2007}, which has been succesfully used for the study of Galactic GCs in regions of the Milky Way with a high density of fore/background stars. Typically, that method considers the colors and magnitudes of the stars in the region under study and compares the distribution in the same plane for a sample of control field stars. In our case, we propose to not only consider the photometry provided by {\it Gaia} DR2 but also the proper motions measured for all the objects in this area. Our final goal will be to estimate the number of probable cluster stars beyond their King and Jacobi radii and to identify any structure they might reveal. A brief description of the followed methodology is described below.

We first select an area around NGC\,362 of 6\,deg $\times$ 6\,deg where we will try to identify the extratidal overdensities. The content of nearby stars in our initial catalog is reduced by  excluding those objects with larger parallaxes ($\varpi > 0.25$\,mas;\, $d_{\odot} < 4$\,kpc). As control objects we use the region within $r = 4.8$\,deg but oustide the studied region, which is equivalent in total area. Based on the CMD and the $\mu$ distribution shown in \figref{figura1}, we only consider stars with $-0.3 < G_{BP_{\rm 0}}-G_{RP_{\rm 0}} < 1.8$, $ 12 < G_{\rm 0} < 21$, $\mu_{\rm 362}-2\sigma_{\mu_{\rm 362}} < \mu < \mu_{\rm 362} + 2\sigma_{\mu_{\rm 362}}$, where $\mu_{\rm 362}$ and $\sigma_{\mu_{\rm 362}}$ are the mean value of the distribution and its standard deviation, respectively. The ($G_{0},G_{BP_{\rm 0}}-G_{RP_{\rm 0}},\mu_{\alpha^{*}},\mu_{\delta}$) space is divided into a grid of cells, where the cell size $\epsilon$ is previously defined. The selected values for this work after testing different combinations are $\epsilon_{\rm G_{\rm BP_{0}}-G_{\rm RP_{0}}} = 0.2, \epsilon_{\rm G_{0}} = 0.5, \epsilon_{\mu_{\alpha^{*}}} = 0.5$ and $\epsilon_{\mu_{\delta}}= 0.5$.  

We then compute the weight ($w$) for all the stars placed in those cells by using the expression

\begin{equation}
w = 1 - \frac{N_{\rm field}\,A_{\rm cluster}}{N_{\rm cluster}\,A_{\rm field}}
\end{equation}

where $N$ and $A$ correspond to the number of stars in a given ($G_{0},G_{\rm BP_{\rm 0}}-G_{\rm RP_{\rm 0}},\mu_{\alpha^{*}},\mu_{\delta}$) cell and the total area for each population (cluster or control field), respectively. When we repeat this procedure over all the cells, we obtain the weights for all the  stars in the  considered region. It is important to avoid any effect that our selection of the cell sizes may have in the results. For this purpose, we repeat the process described above by shifting the grid in the ($G_{0},G_{\rm BP_{\rm 0}}-G_{\rm RP_{\rm 0}},\mu_{\alpha^{*}},\mu_{\delta}$) space in steps of 1/3\,$\epsilon$ in each dimension, yielding 81 different configurations and weight values. We finish this process by assigning the mean value $\overline{w}$ to each star.

\section{Results and discussion}

  \begin{figure*}
     \begin{center}
      \includegraphics[scale=0.6]{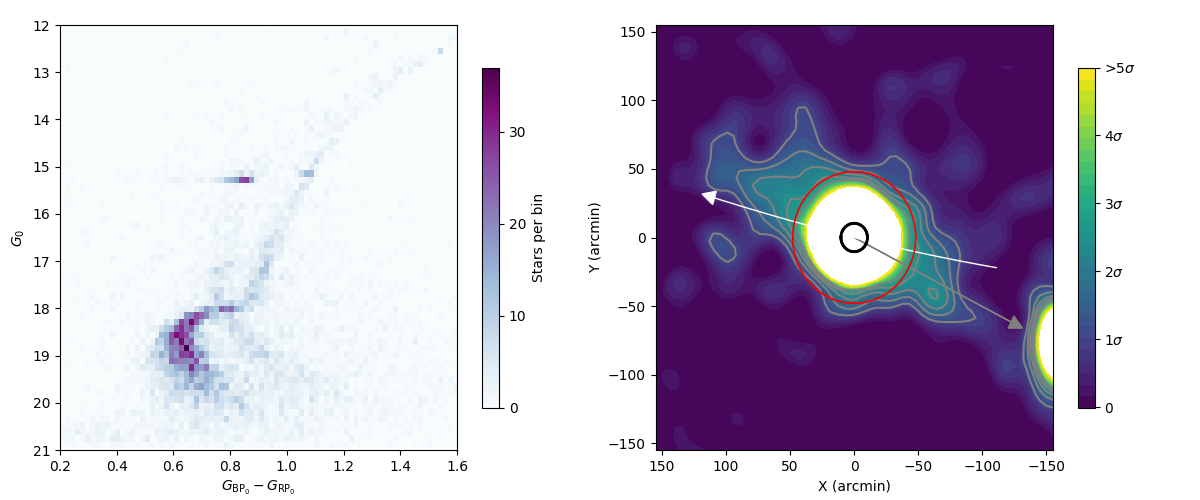}
      \caption[]{\emph{Left:} weighted Hess diagram corresponding to stars within 100\,arcmin from the center of NGC\,362. The low density component observed around $G_{\rm BP_{0}}-G_{\rm RP_{\rm 0}} \sim 0.8$ may be associated with the presence of NGC\,104 stars in the field. \emph{Right:} density map generated for the surroundings of NGC\,362. The overplotted contours are $S$ =[1,1.5,2,2.5,3] and the color code on the right bar indicates the $S$ value over the background population. Note that all the spatial bins in the map with $S > 5$ are represented by the $S=5$ level. The black and red solid lines indicate the King and Jacobi radii, respectively, and the white arrow represents the computed orbit of the cluster based on its {\it Gaia} proper motions. The grey arrow indicates the direction of the Galactic center.} 
\label{figura2}
     \end{center}
   \end{figure*}

The weighted Hess diagram generated with the stars in the region around NGC\,362 is shown in \figref{figura2}. The features in the diagram associated with the fore/background, which are easily seen in \figref{figura1}, are successfully removed, which was one of the main reasons for using this technique on this cluster. When we compare the resulting diagram with the one obtained for NGC\,362 before the cleaning process (\figref{figura1}), it is evident that an important fraction of MS stars have disappeared and more importantly in the $G_{\rm 0} > 19$ section. This is due to our selection criteria based on the proper motion distributions derived for the cluster stars (\figref{figura1}). Given that the mean values of $\mu_{\alpha^{*}}$ and $\mu_{\delta}$ are derived from the brightest stars ($G_{\rm 0} < 18$), faint MS stars with higher errors are not considered during the weighting process. In exchange to this loss of cluster members, these candidates are the stars in this region with higher probability of belonging to NGC\,362. There is not a significant change in the results when using a larger control sky area avoiding the SMC.      

Together with the stars associated with our cluster, there is a second MS in the Hess diagram around $G_{\rm BP_{0}}-G_{\rm RP_{\rm 0}} \sim 0.8$ with $G_{\rm 0} > 17$. Those stars seem to be NGC\,104 members which are also present in the field around NGC\,362. Indeed, those stars likely contributed by the neighbour GC reach the inner regions of our target cluster as a background population \citep[they are separated by $\sim$4\,kpc along the line-of-sight;][]{Chen2018}. Recently, \cite{deBoer2019} estimated a Jacobi radius of $r_{\rm j} = 104.4 $\,arcmin for NGC\,104, thus it seems reasonable to find stars still bounded to that cluster nearby NGC\,362. Moreover, both GCs have very similar proper motions so stars in the faint section of the MS of NGC\,104 fall in our selection window in the ($\mu_{\alpha^{*}},\mu_{\delta}$) space (also due to the photometric errors in that section of the CMD).

We sum the individual weights in bins of 5\,arcmin~$\times$~5\,arcmin and generate the density map shown on the right panel of \figref{figura2}. The map was smoothed using a  Gaussian filter with a width equivalent to 2 bins and the result ($S$) is expressed in standard deviations over the mean value in the field, i.e. $S = ({\rm signal} - {\rm <signal>})/{\sigma_{\rm signal}}$. The saturation level considered to generate the map contained in \figref{figura2} is $S = 5$. We also computed the orbit for NGC\,362 using the \cite{Miyamoto1975} potentials based on the proper motions derived by \cite{Gaia2018b}, a radial velocity of $v_{\rm r} = 223.5$\,km\,s$^{-1}$, the heliocentric distance derived by \cite{Chen2018}, a circular speed of $V_{\rm c}(R_{\rm \odot}) = 220$\,km\,s$^{-1}$, and the Solar motion from \cite{Schonrich2010}. The inmediate orbit of NGC\,362, integrated backwards and forwards, is finally overplotted in the map. The tidal and Jacobi radii of this cluster are $r_{\rm t} = 13.2$\,arcmin and $r_{\rm j} = 47.3$\,arcmin, respectively \citep{Harris2010,deBoer2019}. 

The resulting density map suggests that there exist significant overdensities beyond the $r_{\rm j}$ of NGC\,362 at distances up to almost 2\,deg from the cluster center, which is equivalent to $\sim$300\,pc at the distance of NGC\,362. A second significant group of stars, likely associated with NGC\,104, is observed at $> 3$\,deg from the cluster center. Since both overdensities are clearly disconnected in the map, we conclude that the extratidal features revealed here are mainly composed of member stars of NGC\,362 and are not generated by the presence of NGC\,104 members in the studied area. It is also possible to discard that the largest structure observed is the result of significant gradients in the extinction towards this GC, given its intermediate Galactic latitude ($b$ = -46.25$^{\circ}$) and the low $E(B-V)$ levels in that area with $<E(B-V)> = 0.02 \pm 0.01$. On the other hand, the area in the sky occupied by the SMC presents higher reddening (peaks with $E(B-V) > 1$), which may difficult the detection of a proper tail in that direction.

The most significant density contours shown in \figref{figura2} seem to be elongated in the direction of the orbit derived for NGC\,362 and beyond $r_{\rm J}$. This may indicate that the structure of this GC is affected by its interaction with the Galactic tidal field, thus NGC\,362 could have lost stars along its orbit. This GC is moving on a highly eccentric orbit ($e > 0.85$), reaching an apocenter of $R_{G} \sim 12$\,kpc but also performing incursions into the inner Galaxy with a pericenter of $R_{G} < 1$\,kpc \citep{Gaia2018b}. Globulars with high values of $e$ are subject to important changes in the external field, which is one of the processes leading Galactic GCs to their evaporation. Indeed, numerical simulations and observational evidence suggest that the eccentricity of the orbit has a direct impact in the possibility of generating tidal tails, their morphology and orientation \citep[e.g.][]{Montuori2007,Carballo-Bello2012,Kupper2012}, even for dense and massive GCs like NGC\,362 \citep{McLaughlin2005}. There are no previous reports of extratidal structures around this cluster in the literature, even though stars outside the King tidal radius have been detected \citep{Grillmair1995,Anguiano2015,deBoer2019}.

In the \cite{Zinn1993} young/old halo GC scheme, NGC\,362 is classified as a young cluster, thus it is a good candidate to be an extra-Galactic cluster accreted by the Milky Way (although reading its age from its location in the [Fe/H] vs. HB index plane is difficult). Interestingly, together with other 9 Galactic GCs, it has been proposed as a member of the hypothetical GC system of the massive dwarf galaxy, whose impact with the Milky Way created the recently discovered {\it Gaia Sausage/Enceladus} structure \citep{Belokurov2018,Helmi2018,Myeong2018}. Moreover, most of the globulars in that candidate list present extended stellar structures \citep{Kuzma2016,Carballo-Bello2018,Piatti2019}, which may be related with their formation within a less massive galaxy. 

The algorithm described in this Paper is powerful enough to disentangle the stellar population associated with NGC\,362 from that of the surrounding systems, including NGC\,104 and the SMC. However, we still have to face the levels of uncertainty associated with the parameters in the {\it Gaia} DR2. For instance, for stars in our field with magnitudes in the ranges $14 < G_{\rm 0} < 16$, $18 < G_{\rm 0} < 19$ and $20 < G_{\rm 0} < 21$, the mean errors are $<\sigma_{\rm G_{BP_{0}}-G_{RP_{0}}}> \sim$ 0.01, 0.04, 0.2 and $<\sigma_{\mu}> \sim$ 0.1, 0.3, 1.7\,mas\,yr$^{-1}$, respectively. In this scenario, the weight assigned to the faint stars belonging to NGC\,362 in the ($G_{\rm 0},G_{\rm BP_{0}}-G_{\rm RP_{0}},\mu_{\alpha^{*}},\mu_{\delta}$) cell will significantly change between iterations, while brighter members will remain as candidates with larger $\overline{w}$, even when the cell limits are modified. Therefore,  the number of objects with $G_{\rm 0} >$ 18--19 and large $\overline{w}$ values in the final catalog is only a fraction of the real population of NGC\,362 stars observed by {\it Gaia}. Even so, this methodology seems to be useful in the study of those nearby GCs with measured proper motions, which significantly differ from the ones of the fore/background stellar populations.

The known limitations of {\it Gaia} DR2 in dense areas \citep[i.e. an overrepresented population of bright stars in comparison to fainter objects; see][]{Gaia2016} have direct impact in our ability to unveil extratidal features around nearby clusters. These technical constraints will be attenuated as the mission completes the planned number of scans for each object and the final catalogue is generated. The methods as the one described in this work will benefit from the significantly smaller errors in the photometry, proper motions and parallaxes. It will thus be possible to identify individual MS stars several magnitudes below the TO even when their current DR2 measurements make it difficult to distinguish those stars from the control field populations. The expected end-of-mission accuracy for {\it Gaia} will allow us to better map the regions around nearby GCs in the searching for evidence of tidal disruption and confirm the existence of extended stellar structures in their surroundings.

\section{Conclusions}

In this work we have shown that it is possible to use a simple modification of the classical statistical decontamination procedure on {\it Gaia} DR2 to identify extratidal structures around the Galactic GC NGC\,362. Our results suggest that a significant number of stars likely associated with that cluster lie beyond its King and Jacobi radii, proving that this cluster is losing members along its highly eccentric orbit.

Given the uncertainties both in the photometry and proper motion measurements for the faintest stars in NGC\,362, only a fraction of the population of the cluster is recovered with large weights after the procedure. Although, for now, this algorithm may be used only in those clusters whose proper motions significantly differ from that of the fore/background stellar populations, these results suggest that the {\it Gaia} end-of-mission data will allow us to unveil low surface-brightness structures in the surroundings of Galactic GCs.

\section*{Acknowledgements}

Thanks to the anonymous referee for her/his helpful comments and suggestions. JAC-B acknowledges financial support to CAS-CONICYT 17003 and FONDECYT postdoctoral fellowship 3160502. Thanks to M. Catelan, J. M. Corral-Santana and C. Navarrete for their careful reading. This work has made use of data from the European Space Agency (ESA) mission {\it Gaia} (\url{https://www.cosmos.esa.int/gaia}), processed by the {\it Gaia} Data Processing and Analysis Consortium (DPAC, \url{https://www.cosmos.esa.int/web/gaia/dpac/consortium}). Funding for the DPC has been provided by national institutions, in particular the institutions participating in the {\it Gaia} Multilateral Agreement.

\def\jnl@style{\it}                       
\def\mnref@jnl#1{{\jnl@style#1}}          
\def\aj{\mnref@jnl{AJ}}                   
\def\apj{\mnref@jnl{ApJ}}                 
\def\aap{\mnref@jnl{A\&A}}                
\def\apjl{\mnref@jnl{ApJL}}               
\def\mnras{\mnref@jnl{MNRAS}}             
\def\nat{\mnref@jnl{Nat.}}                
\def\iaucirc{\mnref@jnl{IAU~Circ.}}       
\def\atel{\mnref@jnl{ATel}}               
\def\iausymp{\mnref@jnl{IAU~Symp.}}       
\def\pasp{\mnref@jnl{PASP}}               
\def\araa{\mnref@jnl{ARA\&A}}             
\def\apjs{\mnref@jnl{ApJS}}               
\def\aapr{\mnref@jnl{A\&A Rev.}}          

\bibliographystyle{mn2e}
\bibliography{biblio}

\bsp	
\label{lastpage}
\end{document}